# Probing the surface of synthetic opals with the vanadyl-containing crude oil by using EPR and ENDOR techniques

M. Gafurov, A. Galukhin, Y. Osin, F. Murzakhanov, I. Gracheva, G. Mamin, S.B. Orlinskii*

Kazan Federal University, Kremlevskaya 18, Kazan 420008, Russia
*E-mail: sergei.orlinskii@kpfu.ru*

Porous silica materials offer wide range of possibilities for enhancement of the productivity of oil reservoirs. However the mechanism of adsorption of polar components of crude oil on silica surface is poorly understood that hinders technological improvement of supports and oil extraction. We have synthesized opal films with the silica microspheres size of about $360 \pm 20$ nm, specific surface area of $5.2 \text{ m}^2 \times \text{g}^{-1}$ and pore size of 230 nm. We have fractionated and characterized oil and oil asphaltenes from heavy (Ashalchinskoe) oil. By pulsed electron paramagnetic resonance (EPR) and double electron-nuclear resonance (ENDOR) in the W-band frequency range (microwave frequency of 94 GHz, magnetic field of 3.4 T) we have studied the adsorption of oil asphaltenes on the surface of opal samples using the intrinsic for asphaltenes paramagnetic vanadylporphyrins (VO) complexes. $^1$H ENDOR spectra are found to be different for initial and the adsorbed samples in their central parts (that is a sign of asphaltenes VO disaggregation) whereas no significant changes in the W-band EPR spectra were detected. Contrasting to alumina support, no strong electron-proton interaction with the protons on the surface of $SiO_2$ (presumably, silanol groups) was found and infiltration of the oiled opal films with gasoline changes the central part of $^1$H ENDOR spectra. It shows that the proton containing groups on the surface of amorphous $SiO_2$ sample can significantly change the asphaltene adsorption properties and ENDOR of the intrinsic for oil paramagnetic centers could be used for the characterization of surface state in porous media *in situ* or *operando*.

## 1. Introduction

Mesoporous and nanoporous materials seem to offer limitless possibilities in a wide range of applications, such as catalysis, absorption and adsorption, sensors, optical and photovoltaic devices, fuel cells as well as for biochemical technology like drug delivery or molecular sensing, and still, they have not revealed their entire potential yet. Historically, ordered mesoporous silicas were the first reported class of such materials with regular structures, along with very high specific surface areas, thermal and mechanical stability, highly uniform pore distribution and tunable pore size, high adsorption capacity and unprecedented hosting properties [1-3]. Synthetic opal composed of silica spheres has received great attention last decades also as a template to load with another material and, as a result, to obtain the structure with a complete photonic band gap, for example [4, 5].

Compared with other silica minerals and silica glass, natural opal contains a large amount of water (up to 10 wt. %). Infrared (IR) studies at ambient conditions indicate that protons in these systems are presented by various forms, e.g. free water protons, protons of water adsorbed on the silica surface, and protons of silanol groups (Si–OH) located on the surface of the silica spherules [6]. The existence of silanol groups was proved by NMR [7]. It is known that surface chemistry is a critical factor for determining the behavior of a nanomaterial after incorporation in composites and surface modification greatly influences the functionality and properties of nanosized materials [8-10]. Therefore, analysis and study on the surface of nanoparticles, including magnetic resonance methods, is a key to obtain their important physicochemical properties for the subsequent applications.

Recently, nanoparticles have been presented as alternatives for *in situ* enhancement of oil reservoir productivity [11]. Due to their high surface area/volume ratios and excellent dispersability, silica nanoparticles are able to quickly capture and remove the asphaltenes (the heaviest oil components) which flocculation and precipitation decrease extraction capacity of light oil fractions causing blockages in pipes and pieces of equipment, preventing extraction of lighter hydrocarbons by a simple way and generating losses of energy resources. It has a direct impact on the enhancement of oil production, inhibiting formation damage and restoring wettability. It was shown, for example, that treatment with silica nanoparticles with the higher acidity increased the effective permeability to oil and enhanced the oil recovery on 11 % comparing to the untreated silica surface [11]. Also the influence of nanoconfinement on reactivity of heavy crude oil in combustion process was recently studied [12]. It turned out combustion of oil in nanoporous medium of synthetic opals proceeds faster than in a bulk. This approach has the major benefit of being scalable to a producing field. But the understanding of the roles of size, surface treatment of silica nanoparticles in the wettability alteration and inhibition of formation damage caused by asphaltenes and their influences on asphaltene aggregate size in the oil matrix and the adsorbed phases is still poor investigated.

In our recently published research [13] we studied the adsorption of asphaltenes from heavy oil on the surface of γ-aluminum oxide by electron paramagnetic resonance (EPR) and double electron-nuclear resonance (ENDOR) in the W-band frequency range (microwave frequency of 95 GHz, magnetic field of 3.4 T) using previous experience in alumina [14, 15] and oil fractions [12, 13] characterization. Here we use infiltration of the opal pores with the Ashalchinskoe crude oil containing native vanadyl (VO) paramagnetic centers to probe the interaction with the silica surface with aim to shed light on details of adsorption of oil asphaltenes on silica. The effectiveness of the use of paramagnetic vanadyl complexes as spin probes is determined by the resolved anisotropic spectral parameters (*g*-factor, hyperfine splitiing, zero-field splitting, quadropole tensors) in the EPR/ENDOR spectra [16], the known affinity of porphyrin complexes and oil asphaltenes to silica surfaces [17-21].

## 2. Materials and Methods

Preparation of silica spheres by two-step controllable Stöber growth technique based on regrowth of silica seeds was described in details in paper [22]. Opal films were prepared by modified vertical deposition method based on isothermal heating evaporation-induced self-assembly [23, 24]. Obtained opal films were calcined at 1050°C for 12 hours.

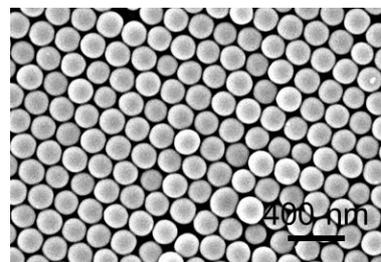

**Figure 1.** SEM image of synthesized opal sample. A scale of 400 nm is shown.

The particle size in colloidal crystals was extracted by using scanning electron microscopy (SEM). SEM measurements were carried out on high-resolution scanning electron microscope Merlin Carl Zeiss (Figure 1). Observation photos of morphology surface were obtained at accelerating voltage of incident electrons of 15 kV and current of 300 pA. The average diameter of the silica spheres was estimated as 360 ± 20 nm.

Nitrogen adsorption and desorption measurements at $T = 77$ K were carried out with ASAP 2020 MP instrument (Micromeritics). Adsorption and desorption isotherms contained about 200 points for each colloidal crystal sample. Specific surface areas of the silica colloid crystal samples were determined to about $5.21 \pm 0.06$ m$^2 \times$ g$^{-1}$ by applying the Brunauer − Emmett − Teller (BET) equation in a range of 0.05-0.30 $P/P_0$ (Figure 2), where $P$ and $P_0$ are the partial (equilibrium) and saturated pressure of the nitrogen gas at 77 K [25]. The pores size in the synthesized opal sample is estimated to be about 230 nm.

EPR investigations were done by using helium flow cryostats on the X-band ESP 300 (microwave frequency ν ≈ 9.5 GHz) and Bruker Elexsys W-band E680 (microwave frequency ν ≈ 93.5 GHz) spectrometers. We used conventional continuous wave (CW) and two-pulse field swept electron spin echo (FS ESE) measurements for the detection of EPR signal and the primary Hahn-echo (π/2-τ-π-echo), where τ is the interpulse delay time of 240 ns, with initial π/2 and π pulse lengths of 32 and 64 ns, respectively. For the W-band ENDOR experiments we used special double (for nuclei and electron) cavities and Mims pulse sequence (π/2-τ-π/2-*T*-π/2) with an additional radiofrequency (RF) pulse $\pi_{RF}$ = 18 µs inserted between the second and third microwave $\pi/2$ pulses. RF frequency in our setup could be swept in the range of 1-200 MHz [13].

We have chosen Ashalchinskoe oil (Volga-Ural basin, Republic Tatarstan, Russia) as one of the most complex petroleum systems but, from other side, as the most studied by different analytical methods including NMR, EPR, X-band ENDOR techniques (see [26-32] and references therein). Mn-catalyzed oxidation of the Ashalcha heavy oil in porous media was examined in ref. [33] by combining X-ray powder diffraction, thermal analysis, non-isothermal kinetic methods and EPR approaches. The total amount of vanadium in Ashalcha oil is estimated to be about 0.023 wt. % distributed mainly in asphaltenes (0.186 wt. %) and resins (0.046 wt. %)

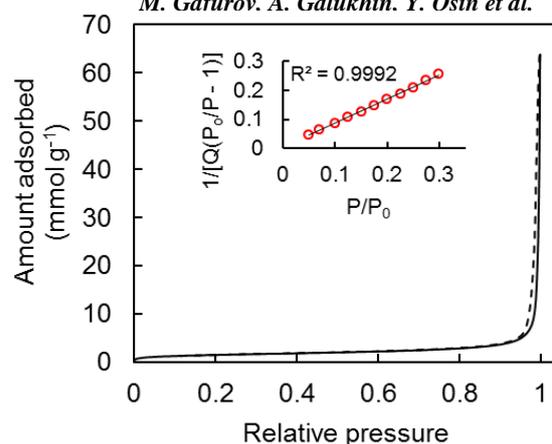

**Figure 2.** Nitrogen adsorption (solid line) and desorption (dashed line) isotherms at $T$ = 77 K and Brunauer – Emmett – Teller plot (inset) of synthetic opal. $R^2$ is the correlation coefficient, $Q$ is an amount of the nitrogen needed for complete monolayer coverage.

[34]. Concentration of the vanadyl paramagnetic complexes was estimated as $1.1 \cdot 10^{18}$ spin/g in the initial oil sample and of $9 \cdot 10^{18}$ spin/g in the extracted asphaltenes from the CW X-band measurements.

The four components of the oil: saturates, aromatics, resins, and asphaltenes (SARA) were separated and quantified using an ASTM method D2007, based on extraction (n-alkane) and adsorption on alumina [35]. The measurements of the dynamic viscosity η were performed with Stabinger viscometer SVM 3000/G2. The extracted data are presented in Table 1.

Opal samples with adsorbed asphaltenes were prepared by soaking of 50 mg of opal in 5 mL of asphaltene solution in dichloromethane with concentration of 2.24 g L$^{-1}$ for 36 hours at 35°C. As an additional source of protons that has a large affinity to the asphaltene solution but does not solve the asphaltene complexes and does not influence the electromagnetic field inside the EPR cavity, Zippo premium lighter fluid (USA), named throughout the work as gasoline, was exploited. In such a manner four types of samples were studied: (1) opal films; (2) oil and oil asphaltenes (denoted further as oil); (3) opal films with the adsorbed oil asphaltenes (denoted as opal + oil) and (4) opal films with the adsorbed asphaltenes placed into the EPR quartz tube filled with gasoline (opal / oil / gasoline).

**Table 1.** Rheological and component analysis data for the Ashalcha oil.

| Viscosity, mPa s | Density, g cm$^{-3}$ | SARA analysis, % | | | |
|---|---|---|---|---|---|
| | | Saturated | Aromatic | Resins | Asphaltenes |
| 2805 | 0.97 | 26.2 ± 0.5 | 43.0 ± 0.6 | 26.3 ± 0.5 | 4.5 ± 0.3 |

## 3. Results and Discussion

No EPR signal was registered in the initial opal samples (1) within the sensitivity limits of our equipment. W-band EPR spectra for the infiltrated species [samples (2)-(4)] are presented in Figure 3.

EPR spectra are very similar to each other and defined by stable "free" radical (FR, single line, electron spin $S = 1/2$, $g \approx 2.003$) and paramagnetic vanadyl complexes. Origin and EPR of FR in oils and oil asphaltenes are comprehensively described in the literature (see [36] and references therein) and are not in the scope of the present study.

Assuming a nearly planar skeleton structure of single VO molecule (VO$^{2+}$, $S = 1/2$, Figure 4), the EPR spectra could be satisfactory described by the $g$-tensor of axial symmetry with $g_{\parallel} \approx 1.96$, $g_{\perp} \approx 1.98$ and anisotropic hyperfine interaction with a magnetic moment $I = 7/2$ for $^{51}$V nuclei (natural abundance of 99.75 %) with hyperfine structure constants $A_\parallel \approx 480$ MHz, $A_\perp \approx 157$ MHz. In our notations the values of $(g, A)_\parallel$ correspond to the orientation perpendicular to $VO^{2+}$ plane (out of plane), along the direction $c$ and $(g, A)_\perp$ correspond to the orientation in the $VO^{2+}$ plane, *ab* plane. Comprehensive description of VO spectra in X and W-bands, their spectra simulations and spectra changes with external treatment are given in papers [26, 37, 38].

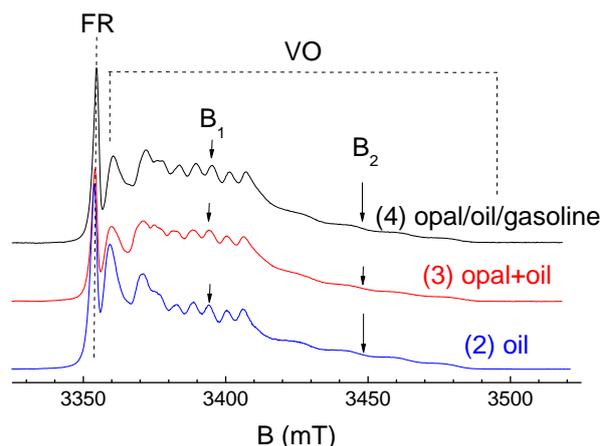

**Figure 3.** W-band FS ESE spectra at $T = 20$ K for the investigated species. Positions of FR line and VO spectra are marked by dashed lines. Values of the magnetic fields $B_1$ and $B_2$ in which $^1$H ENDOR spectra were registered are shown.

While there is no significant changes in EPR spectra were detected, a significant difference in the central part of the spectra of the proton ENDOR with adsorption is obtained (Figure 5). In ref. [26] the reason of choice of the external magnetic fields denoted as $B_1$ and $B_2$, correspondingly, in which the ENDOR spectra were acquired, was given. Briefly, for the convenience such values were chosen that correspond to "pure" parallel and perpendicular orientations of the external magnetic field relative to VO geometry: $B_1$ value corresponds to $B_0$ "out-of-plane", along *c*-axis while $B_2$ for that "in-plane" for $m_I = 3/2$ transitions. Besides of being from the same $m_I$ (that simplifies the interpretation and theoretical analyses), the next reasons play a role: (1) the sufficient ESE amplitudes to obtain a reasonable signal-to-noise ratio; (2) absence of overlapping with the shoulders of the FR signal and (3) absence of the orthogonal contributions.

$^1$H ENDOR pattern (splittings along the ENDOR spectrum) for asphaltene and oil sample are different for the parallel and perpendicular orientations, are in the range of 0.3-2.4 MHz and could be ascribed to the interaction with the protons of the following structures of VO cores identified in natural hydrocarbons: etioporphyrin (VOEtio), deoxophylloerythroetioporphyrin (VODPEP), and benzoetioporphyrin (VOBenzo, see paper [26]). According to the performed density functional theory (DFT) calculations, they correspond to the interatomic distances between the vanadium ion and the closest protons of the VO cores (Figure 4) in the range of 0.46-0.6 nm.

Comparing to asphaltenes, in the adsorbed samples the only central part is changing: the central peak, which is due to the hyperfine interaction of asphaltene VO complexes with a large number of the remote protons, disappears signing the disaggregation of the VO complexes. From the linewidth of the symmetrical narrow peak (in assumption of the simple electron-proton point-dipole approximation which is applicable for the large distances between electron and nuclear spins) the distance can be estimated as ~ 0.8 nm and larger between the paramagnetic centers and the protons, and is observed for both (parallel and perpendicular) orientations. This value is comparable with the size of asphaltenes

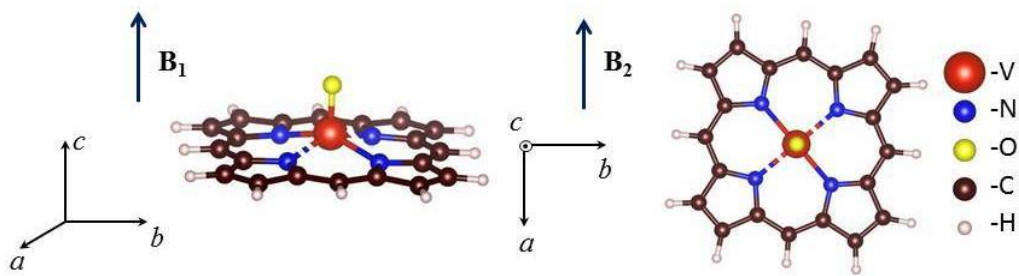

**Figure 4.** Schematic representation of a skeleton of single vanadyl porphyrin molecule. The directions of the external magnetic fields "in plane" ($B_2$) and "out of plane" ($B_1$) for the W-band ENDOR measurements are shown (see Figure 3).

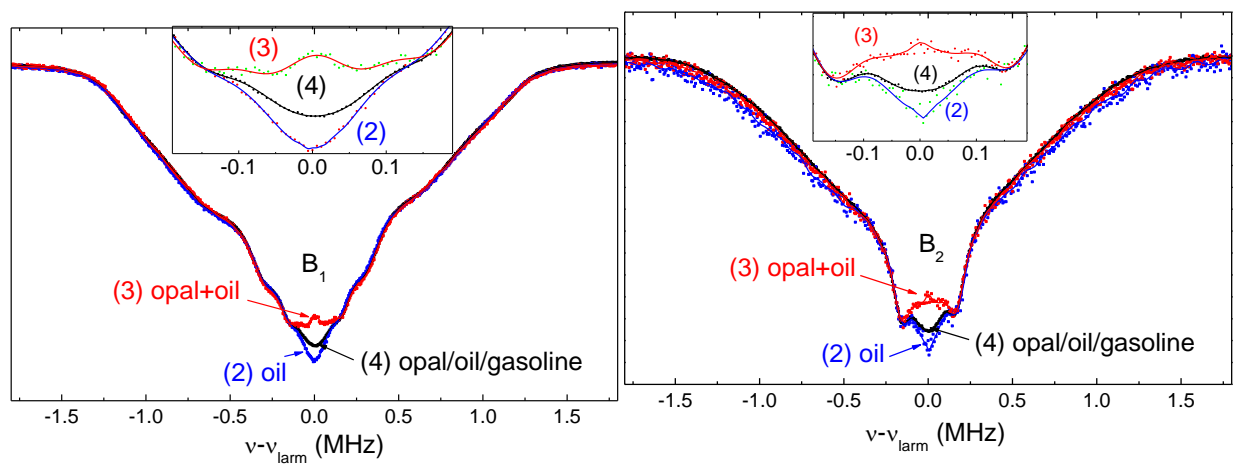

**Figure 5.** $^1$H ENDOR spectra near the proton Larmor frequency measured for oil and oil infiltrated opal samples (2) - (4) at $B_1 = 3395.3$ mT (upper panel) and $B_2 = 3448.9$ mT (lower panel). W-band, $T = 20$ K.
of 1-2 nm reported in literature [39].

It is common knowledge that intermolecular interactions, such as hydrogen bonding, dipole-dipole interactions, π-π stacking etc. in asphaltenes lead to aggregation into nanoaggregates in the bulk oil phase although the basic asphaltene structure and mechanism of aggregation continue to be debated. The relationship between petroporphyrins and asphaltenes is largely unexplored [39]. Dickie and Yen were the first who proposed an aggregation scheme for the macrostructure of petroleum asphaltenes, within which petroporphyrins could stack to the aromatic sheets of the asphaltene [40]. Based on the ENDOR results, mainly on difference between the $^1$H ENDOR spectra for $B_1$ and $B_2$ obtained by adsorption of Karmalinskoe oil asphaltenes on $Al_2O_3$ surface [13], it was concluded that the association of the vanadium petroporphyrins in asphaltene fraction is due to functionality appended to the porphyrin ring, rather than favorable π-π stacking interactions of aromatic rings with the porphyrin core itself. Indeed, for $B_0 = B_2$ additional ENDOR lines with the splittings of 1.3 and 2.2 MHz appeared and $^{27}$Al ENDOR spectra with the strong interaction value (of 1 MHz) was observed on alumina. Soaking of the adsorbed on alumina samples in gasoline had no effects on ENDOR spectra [13].

Contrasting to the results obtained on alumina [13], no additional ENDOR lines with the splittings of 1.3 and 2.2 MHz were registered at $B_0 = B_2$ in the presented opal / oil ENDOR experiments, i.e. ENDOR spectra are orientation independent, and soaking in gasoline leads to the "return" of the central ENDOR peak (Figure 5). The obtained feature could be ascribed to the weak interaction with the surface silanol groups and gasoline protons meaning that the asphaltene adsorption on the investigated silica surface is less than this for alumina.

Our data are in apparent contradiction with the results of Franco et al. [41] who examined the effects of different nanoparticles using asphaltene adsorption and displacement tests in porous media under reservoir conditions. They found that asphaltene adsorption decreased in the following order: silica gel (amorphous) > silica gel (crystalline) > alumina > washed rock. It emphasizes the necessity of proper surface treatment and applying various analytical tools for its characterization.

## 4. Conclusion

In this work we have studied the adsorption of oil asphaltenes from Ashalchnskoe oilfield on the surface of the synthesized opal samples using the intrinsic for asphaltenes paramagnetic petroporphyrin (vanadyl) complexes. $^1$H ENDOR spectra are found to be different for initial and the adsorbed samples in their central parts while no significant changes in the W-band EPR spectra was detected. Contrasting to alumina support [13], no strong electron-proton interaction with the protons on the surface of $SiO_2$ (presumably, silanol groups) was found and infiltration of the oiled opal films with gasoline changes the central part of $^1$H ENDOR spectra. It shows that the proton containing groups on the surface of amorphous $SiO_2$ sample can significantly change the asphaltene adsorption properties.

Conventionally, asphaltenes are regarded as the most annoying component of oil systems responsible for the oil high viscosity, clogging the oil transport, catalyst degradation and deactivation in hydrotreatment units, etc. High stability of asphaltenes and asphaltene metal complexes (including vanadyl porphyrins) hinder the usage of oil as an energy source, as a source of metals [42], as elements of photovoltaics [43], etc. It causes the attention to their studies and beneficial use as soon as the supply of the conventional oil resources continues to decline. Our investigations show that as a profit naturally containing, inevitable for the most heavy oils and bitumen stable vanadyl complexes of asphaltenes can be exploited as a convenient alternative to the known commercial (and often expansive) spin probes at least for the investigations of catalytic surfaces due to the nearly planar structure and unique (orientation dependent) EPR parameters. We hope that usage of vanadyl containing asphaltenes could shed light on the nature of active sites on amorphous silica surface, to study the processes of chemical modifications of the opal surface in the micro and nanoscale porous systems and potentially follow the processes of chemical reactions in the confined media.

**Acknowledgments**

This research was conducted with support of the Russian Science Foundation (Project № 17-73-10350).